\documentclass[fleqn,usenatbib]{mnras}

\usepackage{newtxtext,newtxmath}
\usepackage[T1]{fontenc}

\usepackage{graphicx}	
\usepackage{amsmath}	
\usepackage{xcolor}
\usepackage[flushleft]{threeparttable}
\usepackage{verbatim}
\usepackage{array}
\usepackage{tabularx}
\usepackage{colortbl}
\usepackage{hyperref}
\usepackage{float}
\usepackage{subcaption}
\usepackage{booktabs}
\usepackage{adjustbox}
\usepackage{pdflscape}

\DeclareRobustCommand{\VAN}[3]{#2}
\let\VANthebibliography\thebibliography
\def\thebibliography{\DeclareRobustCommand{\VAN}[3]{##3}\VANthebibliography}

\newcommand{\TESS}{\textit{TESS}}

\title[TESS Positional Probability]{The positional probability and true host star identification of TESS exoplanet candidates}

\author[A. Hadjigeorghiou and D. J. Armstrong]{\parbox{\textwidth}{\Large
Andreas~Hadjigeorghiou$^{1,2}$\thanks{E-mail: A.Hadjigeorghiou@warwick.ac.uk},
David~J.~Armstrong$^{1,2}$}
\vspace{0.2cm}
\\
$^{1}$Department of Physics, University of Warwick, Gibbet Hill Road, Coventry CV4 7AL,\\
$^{2}$Centre for Exoplanets and Habitability, University of Warwick, Gibbet Hill Road, Coventry CV4 7AL, UK}

\date{\vspace{-0.5cm}Accepted XXX. Received YYY; in original form ZZZ}

\pubyear{2023}

\begin{document}
\label{firstpage}
\pagerange{\pageref{firstpage}--\pageref{lastpage}}
\maketitle


\begin{abstract}
We present a method for deriving a probabilistic estimate of the true source of a detected \TESS\ transiting event. Our method relies on comparing the observed photometric centroid offset for the target star with models of the offset that would occur if the event was either on the target or any of the \textit{Gaia} identified nearby sources. The comparison is done probabilistically, allowing us to incorporate the uncertainties of the observed and modelled offsets in our result. The method was developed for \TESS\ Full Frame Image lightcurves produced from the \textit{SPOC} pipeline, but could be easily adapted to lightcurves from other sources. We applied the method on 3226 \TESS\ Objects of Interest (TOIs), with a released lightcurve from \textit{SPOC}. The method correctly identified 96.5\% of 655 known exoplanet hosts as the most likely source of the eclipse. For 142 confirmed Nearby Eclipsing Binaries (NEBs) and Nearby Planet Candidates (NPCs), a nearby source was found to be the most likely in  96.5\% of the cases. For 40 NEBs and NPCs where the true source is known, it was correctly designated as the most likely in 38 of those. Finally, for 2365 active planet candidates, the method suggests that 2072 are most likely on-target and 293 on a nearby source. The method forms a part of an in-development vetting and validation pipeline, called \texttt{RAVEN}, and is released as a standalone tool. 
\end{abstract}




\section{Introduction} \label{sec:intro}
Large scale transit surveys have revolutionised the field of exoplanet science. Ever since the launch of the \textit{Kepler} mission \citep{Kepler}, the number of confirmed and potential exoplanets has rapidly increased. This trend has been further reinforced by the success of the \textit{TESS} mission \citep{TESS}, the first space based, all-sky transit survey, launched in 2018.   

However, transit surveys have an inherent issue in the form of aperture light blending \citep{Brown2003}. The effect is caused by light from background or foreground stars been captured in a target's aperture. Despite the efforts of mission pipelines, such as the \TESS\ Science Processing Operations Center (\textit{SPOC}) \citep{SPOC}, to produce the most optimal aperture, some light from those nearby sources will unavoidably be present in the observed flux. In order to minimise the issue, the \textit{Kepler} mission pipeline implemented a correction on the observed flux to account for these additional contributions \citep{Correction, KeplerPDC}. This was subsequently carried over into the \textit{TESS} pipeline. 

Even with the above correction, a significant problem caused by light blending still remains. In the event that a nearby star hosts an eclipsing companion and contributes enough flux into the target's aperture, the eclipsing event can be detected in the observed flux \citep{Centroid}. Due to the contributing light being only a fraction of the observed flux, the depth of the event can appear much smaller, leading to a False Positive (FP) detection of a planetary companion on the target. This is much more commonly caused by Nearby Eclipsing Binaries (NEBs), although Nearby Transiting Planets (NTPs) have also been observed. 

Identifying the true host of a detected event is therefore an important step in the process of confirming exoplanet candidates. To that effect, extensive follow-up observations are made, covering both the target and any nearby hosts. Part of their purpose is to test whether the detected event is present with a larger eclipse depth on any of the nearby sources, which would indicate that the event is on that source. To assist with the identification, especially due to the large number of exoplanet candidates, several algorithmic methods were developed to identify or provide probability estimates of the true source through the existing or additional observations \citep{BLENDER, Centroid, pastis, DAVE, Triceratops}. 

One such method is the Imaging Difference described in \citet{Centroid}. This method compares averaged target pixel values of in and out of transit frames, to find the pixel position with the highest change in the measured flux. In the event that the only change during the transits is the flux drop due to the eclipse, which is most often the case, then this pixel would correspond to the location of the transit. A Pixel Response Function (PRF) model fit can then be performed to reveal the sub-pixel centroid location of the event. This is a powerful method for the identification of the true host, as it can possibly directly reveal its location. It is is performed automatically by the \textit{SPOC} pipeline once a candidate has been found and is also used in the \TESS\ Quick Look Pipeline (QLP) \citep{QLP1, QLP2} and the exoplanet vetting pipeline \texttt{DAVE} \citep{DAVE}. These often adopt a 3$\sigma$ significance difference threshold between the target location and the transit location to flag the event as off-target. In an effort to improve upon the threshold approach, the Imaging Difference method was expanded in \citet{Bryson2017}, which introduced a probabilistic approach in identifying the true host of the transit. This method utilises generative modelling of the imaging difference for the known sources around the target to derive the probability of the source being the host of the event.

An alternative method, also described in \citet{Centroid}, involves tracking the mean position of the "centre of light" in a collection of pixels, called the photometric centroid. These pixels are usually the aperture pixels for \textit{TESS}. In this method, the mean flux-weighted horizontal and vertical pixel position is computed, which represents the position of the target's photosphere in the aperture. The observed centroid position for cadences in and out of the eclipse event are compared, to reveal any potential offsets correlated with the event. The basis for this comparison is that the loss of light during the eclipse will have an effect on the mean position of the flux. Ideally, if the eclipse is occurring on the target it will result in the centroid remaining unchanged. However, this requires the target to be fully aligned with the "centre of light" in the pixels. For diluted apertures, this will not be the case and an offset is normally observed. Therefore, it is challenging to distinguish on-target and off-target events. The \citet{Centroid} paper provides a detailed description of how the offset can be used to identify the location of the event. Similar to the Imaging Difference technique, this method assumes that the only flux change during the transit is the event itself and is thus susceptible to other sources of photometric variability.

Finally, statistical validation pipelines such as \texttt{PASTIS} \citep{pastis, pastis2}, \texttt{vespa} \citep{vespa} and \texttt{triceratops} \citep{Triceratops}, can also be used to obtain further insight into the source of the event. These pipelines aim to probabilistically confirm the planetary nature of exoplanet candidates, especially for those that are not amenable to follow-up observations. While their implementations differ, all pipelines focus on applying models on the transit lightcurve based on different planetary and astrophysical FP configurations. This allows them to calculate the probability that the observed event belongs to different FP scenarios and then combining them into one final FP probability. If the overall FP probability is sufficiently small, the candidate is then validated. Blended host sources are included among the FP scenarios, although only \texttt{triceratops} examines the actual population of the target's nearby sources, with \texttt{vespa} and \texttt{PASTIS} using a simulated population instead. However, none of the existing validation methods can currently determine the true host of the transit as they do not incorporate any observational positional information for the event.

In this work, we present a method for providing a probability estimate on the true host of a detected \TESS\ event, based on the photometric centroid offset described above. The method is designed to run fast, on a large scale and with minimal additional resources other than the \TESS\ \textit{SPOC} lightcurves. 
It should be noted that the method does not replace the need for follow-up observations to identify with certainty the location or the true host of the event. It does not statistically confirm the location of the event nor does it definitively label events as on or off target based on probability thresholds, as that is not its purpose. Instead, it aims to provide a framework for examining the observed transiting event and deriving a probability estimate for the likelihood of the event being on any of the possible sources, including the target. Testing the method on known on-target and off-target events revealed that it can produce useful probabilities for the majority of cases and offer a strong indication as to the true source of the event. 

A detailed breakdown of the methodology is presented in Section \ref{sec:Method}. The testing of the method on known on-target and off-target \TESS\ Objects of Interest (TOIs), is showcased in Section \ref{sec:Testing}, with the discussion on its effectiveness and limitations in Section \ref{sec:Discussion}. Concluding remarks can be found in Section \ref{sec:Conclusion}. The method was developed as part of a new vetting and validation pipeline for \TESS\, called \texttt{RAVEN} (RAnking and Validation of ExoplaNets), which will be presented in a future upcoming paper. It is released here as a standalone tool, with the code made available on GitHub\footnote{https://github.com/ahadjigeorghiou/TESSPositionalProbability}.

\section{Methodology} \label{sec:Method}
\subsection{Overview}\label{Method:Overview}
The main purpose of the method is to establish a probabilistic framework for examining observed transiting events and their associated photometric centroid offsets, capable of providing probability estimates on the true host stars of the event. 
To do so, the method first takes as input a list of \TESS\ candidates and their transit characteristics, namely the epoch, period, transit duration and depth. Then, for each candidate, the method employs the following steps:

\begin{enumerate}
  \item Determine the magnitude and direction of observed centroid offsets during the transits. 
  \item Identify all nearby stars within a 168$''$ radius from the target, by querying the \TESS\ Input Catalog (TIC) \citep{TIC}, which is derived from \textit{Gaia} \citep{GAIA} .
  \item Construct a modelled version of the observed and averaged \TESS\ target pixel array using the \TESS\ PRF models and the \TESS\ magnitude and position of the identified stars.
  \item Calculate the flux fraction contribution of each star, including the target, in the model aperture and determine the implied eclipse depth for the event originating on each of the nearby sources.
  \item Determine the flux-weighted horizontal and vertical centroid positions for the model aperture pixels, in the absence of any transit.
  \item Iteratively scale the flux of each source by the implied depth to reproduce the effect of a transit, constructing a new model observation and determining the resulting in-transit centroid positions.
  \item Produce a modelled centroid offset for each source and probabilistically compare it to the observed, obtaining the likelihood that the transit is located on the source.
  \item Use Bayesian inference to compute the positional probability for each source.
\end{enumerate}
In the case of multiple sectors of observations, the above process is repeated per sector, producing a collection of probabilities. The median of the sector probabilities for each source is then taken. The median probabilities are re-normalised, providing their final positional probability. The above steps are presented in more detail in the rest of this Section.

\subsection{Observed Centroid Offset}\label{Method:Centroid}
\begin{figure*}
    \centering
    \includegraphics[width=\textwidth]{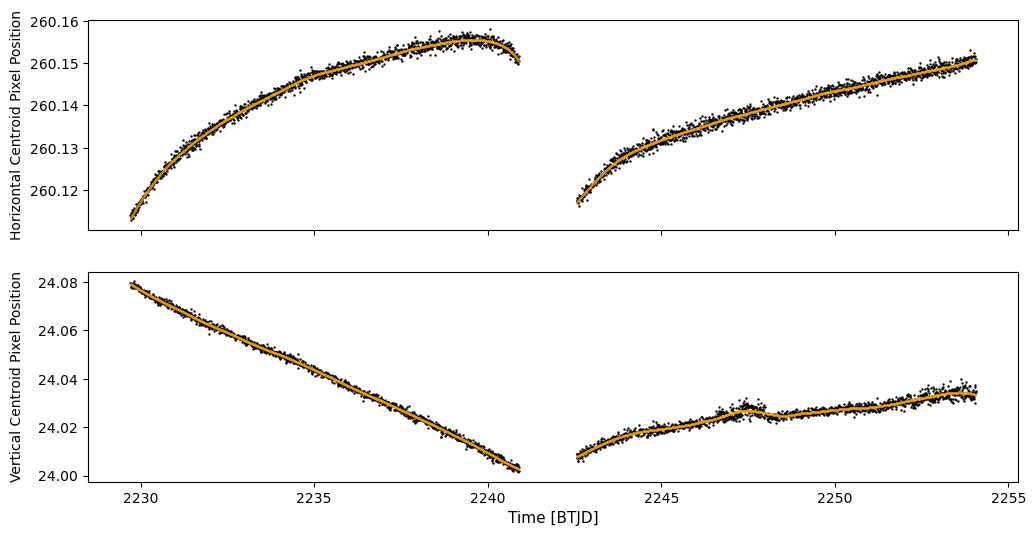}
    \caption{Variation in the horizontal (top) and vertical (bottom) centroid pixel position throughout the observation time for TIC 279740441 in Sector 34, with their corresponding detrending filters overlaid in orange. BTJD = Barycentric corrected \TESS\ Julian
Date.}
    \label{fig:centcurve}
\end{figure*}

The method first calculates the observed photometric centroid offset occuring during the eclipse. As already described in Section \ref{sec:intro}, the photometric centroid is the measured flux-weighted, average horizontal, $C_{x}$, and vertical, $C_{y}$, position of the flux in the aperture throughout the observation. The horizontal position is calculated through the following formula:
\begin{equation} \label{eq: cent}
    C_x = \frac{\sum\limits_{i,j}^{} i f_{ij}}{\sum\limits_{i,j} f_{ij}},
\end{equation}
where i, j are the column and row indices of the aperture pixels respectively, and $f_{ij}$ is the flux contained in each pixel. The vertical position is similarly computed, with the column and row indices in equation \ref{eq: cent} swapped. The centroid offset is then defined as:
\begin{equation} \label{eq: offset}
    \boldsymbol{\Vec{\Delta C}} = (\Delta C_{x}, \Delta C_{y}),
\end{equation}
the vector of the horizontal and vertical pixel position shift of the photometric centroid between the out of transit and in transit cadences. 

For the \textit{SPOC} Full Frame Image (FFI) lightcurves \citep{FFI}, the centroid position per cadence is provided by the pipeline in the MOM\_CENTR1 and MOM\_CENTR2 fields of the lightcurve file, which correspond respectively to the horizontal and vertical centroid position on the CCD. Due to the telescope's motion, the positions exhibit correlated long term variation, as seen in Figure \ref{fig:centcurve} of the $C_{x}$ (top) and $C_{y}$ (bottom) positions for TIC 279740441 in Sector 34. To correct this, the two time-series are detrended, using our own implementation of a Savitzky-Golay filter \citep{SavitzkyGolay}, and normalised. A moving $3^{rd}$ degree polynomial is used for the filter, with a window length of 2 days. The size of the window is much larger than the step size of the window, which is set at 0.15 days, allowing us to guard against overfitting. Data points in transit are masked before detrending to avoid any potential alteration. As the first 12 hours of observation in each sector were found to generally exhibit more extreme variation compared to the rest, these data are removed before detrending. Data points with NaN values are also removed. Following detrending, we remove any outliers with difference greater than 6 times the Median Absolute Deviation from the median, with the process done separately for the in-transit and out of transit data. 

To obtain the magnitude and direction of the centroid offset, the two centroid time series are phasefolded on the period and epoch associated with the transit and fitted with a zero-mean, trapezium transit model. This is due to the offset appearing in the centroid curve similarly to a transit, with the distinct difference that the direction can be either positive or negative and that both the magnitude and direction will vary for each sector. It should be noted that while the fit is performed on the phasefolded centroid curve, the epoch and period parameters are not included in the fit and are used as provided by the user. As such, the method assumes that the provided transit parameters are correct. For the work presented here, the period of the transit was refined as described in Section \ref{Testing-Data}. 

The trapezium model is parameterised by the total transit duration, the full eclipse duration, which excludes the ingress and egress, and the depth. The fit is performed in a robust least-squares framework, based on \texttt{scipy}'s curve\_fit method, using a Trust Region Reflective algorithm \citep{TRF} and a Huber loss function \citep{Huber}. Before fitting the centroid curves, the model is first applied on the transit in the lightcurve itself. The resulting duration parameters are used to initialise the model for the centroid fit, as the offsets are directly related to the transits themselves and thus share the same duration profiles. The transit depth from the fit is stored and used in a later step. It should be noted that the lightcurve is detrended similarly to the centroid curve, with the exception of not removing the first hours of the obervation or outliers.

The centroid fit is then performed. The two duration parameters are allowed to vary by 1\% and 5\% respectively, to account for uncertainties in their determination. The depth is initialised as 0 and can be either positive or negative. The resulting depth from each fit is thus taken as the measured centroid offset in the vertical and horizontal axes, $\Delta C_{x}$ and $\Delta C_{y}$, which form the observed centroid offset vector, $\boldsymbol{\Vec{\Delta C_{obs}}}$. The errors of the two centroid components are calculated from the covariance matrix of the fit. 

\begin{figure}
    \centering
    \includegraphics[width=\columnwidth]{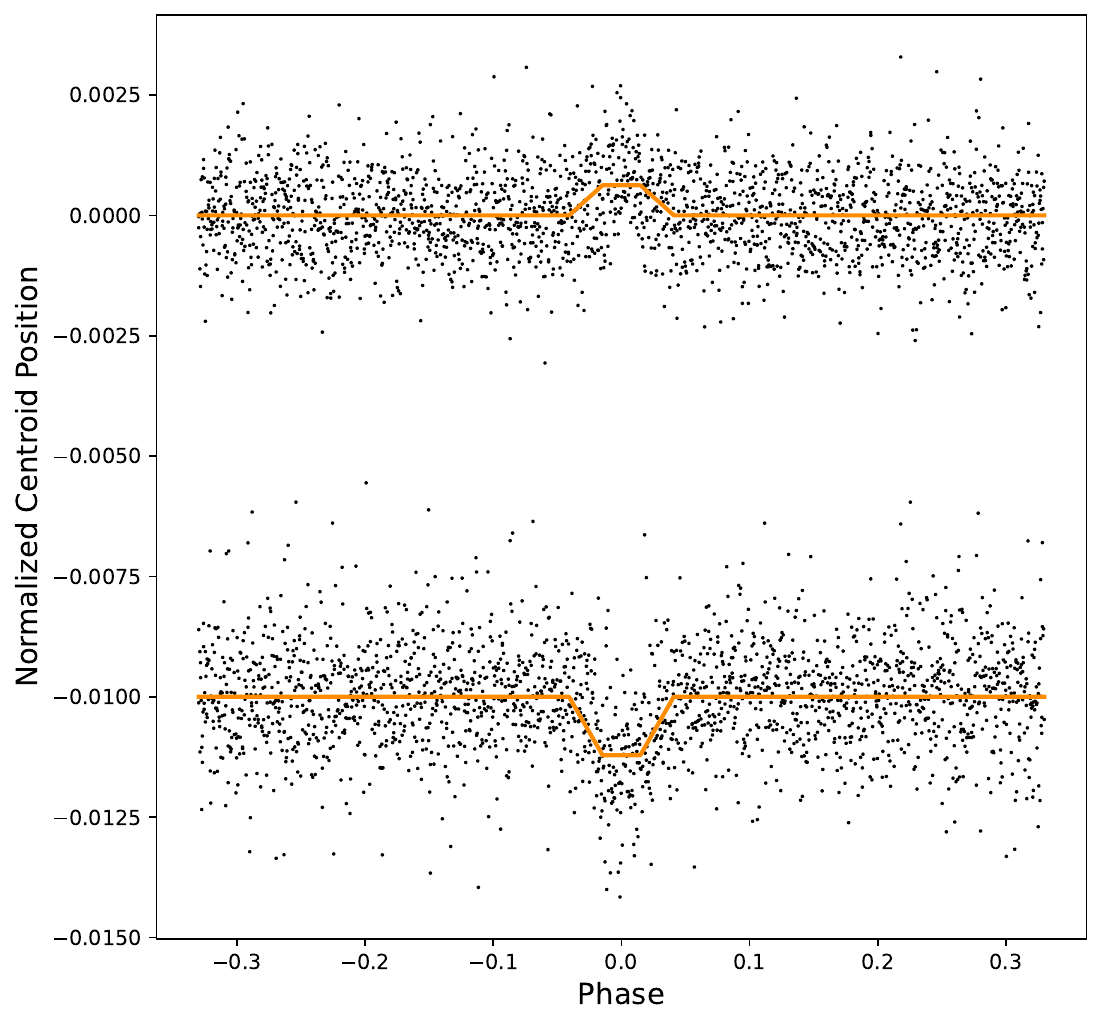}
    \caption{Normalized horizontal (top) and vertical (bottom) centroid position for TIC 279740441 in Sector 34, phasefolded on the period of TOI-273.01. The vertical centroid position has been offseted for clarity. A positive horizontal and a negative vertical centroid offset is clearly seen. The fitted trapezium model for each direction is displayed in orange.}
    \label{fig:centfit}
\end{figure}

A representative example of centroid fitting is shown in Figure \ref{fig:centfit} for the centroid offset associated with TOI-273.01 (TIC 279740441) on Sector 34. Of note, is the fact that a negative offset is seen on the vertical direction and a positive offset on the horizontal direction, suggesting that the potential true source is positioned above and to the left of the target on the CCD. However, it should be stressed that this is not by itself conclusive, as, even if the event is on target, an offset is still expected to be seen.  

Finally, the process described in this section is also used to assess the quality and suitability of the observations for the purpose of our framework. Sector observations are flagged if any of the following criteria is true:
\begin{enumerate}
    \item Less than half of the initial data points remain following the processing and detrending of the centroid curves.
    \item The transit fit for the lightcurves fails or returns a depth less than 50ppm, suggesting that the transit could not be detected in this sector.
    \item The centroid fit for either the horizontal or the vertical axes fails.
\end{enumerate}
The flagged sector observations do not contribute in deriving the positional probabilities for the event. If all observations for an event are flagged, no probabilities will be produced. 

\subsection{Identifying nearby sources}\label{Method:Nearby}
\begin{figure}
    \includegraphics[width=\columnwidth]{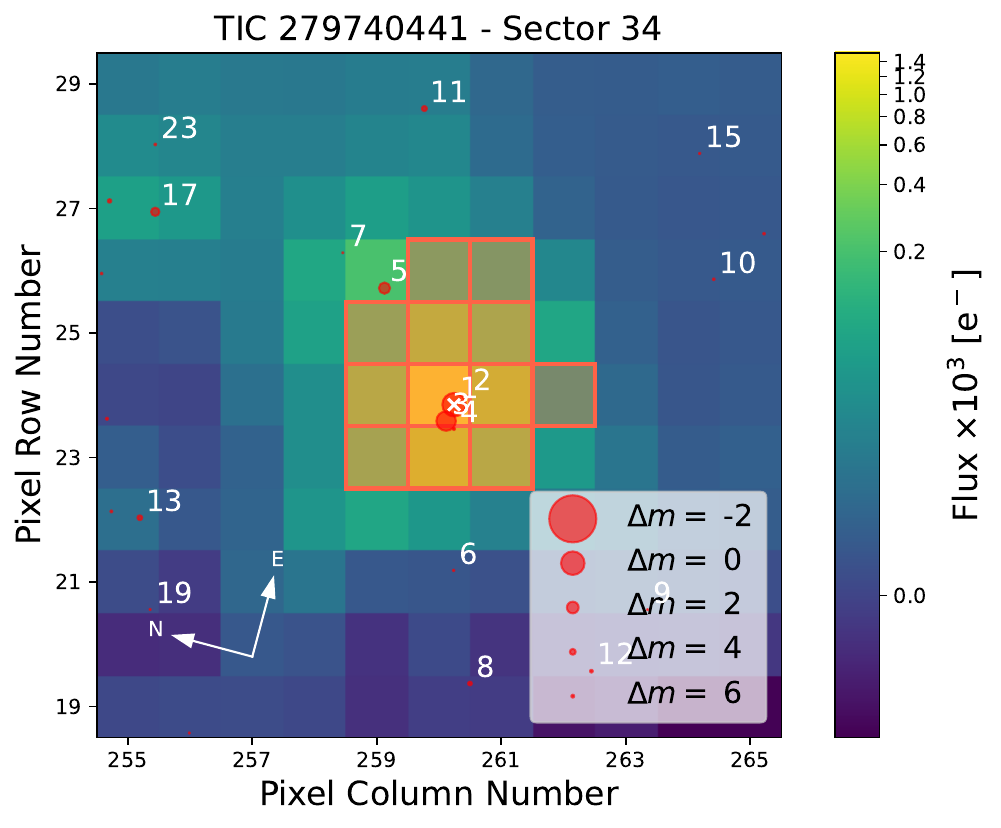}
    \caption{Observed \TESS\ target pixel array for TIC 279740441 in Sector 34. Detected stellar sources are plotted as red circles with their size depending on their magnitude difference with the target. The location of TIC 279740441 is marked with an x. The shaded pixels represent the \textit{SPOC} aperture.}
    \label{fig:TP}
\end{figure}

   The next step is to identify all nearby sources within a $168''$ (8 \TESS\ pixels) from the target, by querying the TIC. The 8.2 version of the catalog was used, which includes flags for "phantom" sources. Sources flagged as "artifacts" or "duplicates" in the TIC are excluded by the algorithm and are not processed further. For all identified sources, their position, proper motion,  \TESS\ magnitude  and object type is retrieved. Due to computational and efficiency reasons, all nearby sources with a \TESS\ magnitude difference greater than 10 are rejected. This threshold stems from computing the magnitude difference between a target and nearby star that would produce a 100ppm drop in the target's lightcurve, assuming that the flux of both the target and the nearby would be fully contained in the aperture, with no other contributing sources. Although these conditions will not hold true for the observations, leading to a smaller magnitude difference, this provides us with an upper limit for any possible nearby host. As an example, the identified nearby sources for TOI-273.01 are overlaid on the target's \TESS\ target pixel array in Figure \ref{fig:TP}, generated through a modified version of \texttt{tpfplotter} \citep{tpfplotter}.
 
 With the nearby sources identified, their \TESS\ magnitude, \textit{Tmag}, is translated into the expected flux count detected with \textit{TESS}, \textit{F}, using the following calibration relation:
 \begin{equation}
    F = A*15000 * 10^{-0.4(Tmag-10)}.
\end{equation}
 This calibration is based on TESS' photometric specification which states an expected flux of 15000 $e^{-}/s$ observed for a 10 magnitude star \citep{TESSHandbook}. The corrector factor \textit{A} is used to adjust the flux for variations in photometry between sectors. It does so by leveraging the fact that the PDCSAP flux for the target in the SPOC lightcurves has been corrected for dilution and as such corresponds to the total flux observed from the target in the sector. The factor is therefore determined by matching the expected flux for the target to the median of the observed flux. The flux calibration step concludes the identification of the nearby sources. Their viability as possible alternative hosts of the observed eclipsing event will then be tested based on two key criteria; the implied eclipsed depth and their suitability to cause the observed centroid offset.

\subsection{Implied eclipse depth for nearby sources}\label{Method:Depth}
To determine the implied eclipse depth, the flux contribution of each source into the target's aperture needs to be quantified. This is done by utilising the TESS PRF models, the position of the identified sources on the TESS pixels during the observation and the expected flux count determined above to effectively re-construct the observation by creating a model target pixel array. Modelling the pixel array allows us to determine the exact amount of flux from each source that is contained within each pixel. We then sum up the flux that falls within the aperture pixels to determine the fraction that each source contributes.

First, the pipeline aperture for the target is retrieved for each sector, along with the \TESS\ PRF models that correspond to the camera and CCD on which the target was observed. In total, 25 PRF models are available per camera and CCD combination, covering different sections of the pixel grid. A PRF model for the observed target pixel array is then produced through interpolation of the 4 closest grid locations. This was implemented based on the publicly released code \texttt{TESS\_PRF} \citep{PRFcode} using the PRF data downloaded from MAST\footnote{https://archive.stsci.edu/missions/tess/models/prf\_fitsfiles/}.

The position of the target and its nearby sources on the pixel array is then determined. For this, their celestial coordinates are updated to match the time of the observation, using the \textit{Gaia} DR2 proper motion listed in the TIC. The coordinates are transformed into pixel positions using the World Coordinate System (WCS) information contained in the \textit{SPOC} lightcurve file and \texttt{astropy}'s WCS module.
The PRF model is then centred on their position, to quantify the distribution of their flux on the pixels. This assumes that any variations to the PRF due to the location of the nearby stars will be negligible. 

\begin{figure}
    \centering
    \includegraphics[width=\columnwidth, trim={0cm 0cm 0cm 0cm}]{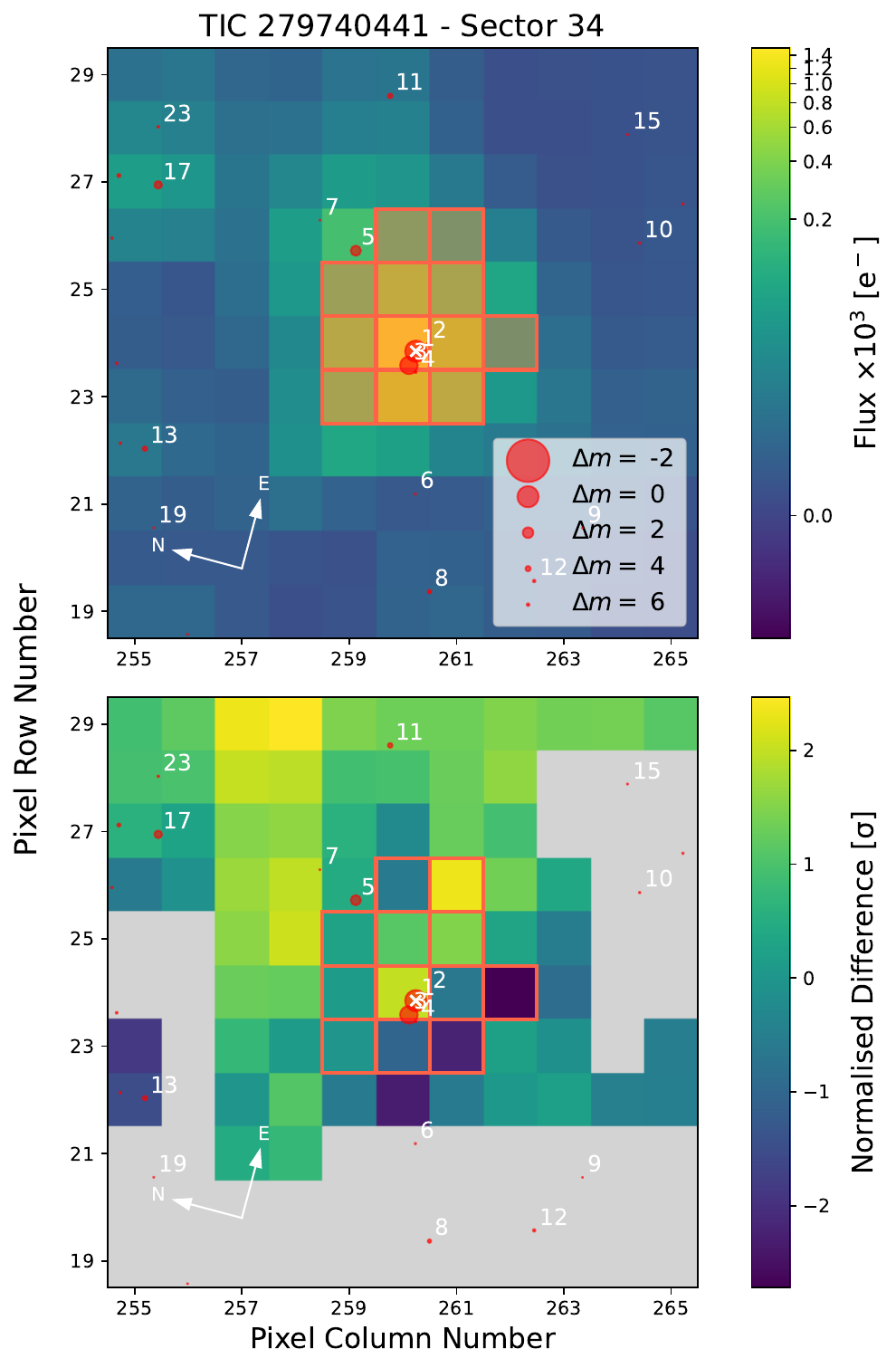}
    \caption{Top: Simulated \TESS\ target pixel array for TIC 279740441 in Sector 34, constructed through the use of the \TESS\ PRF models.
    Bottom: Comparison plot between the simulated and observed pixel arrays, with the flux difference normalised by the shot noise in the observation. Positive values suggest the observed pixel was brighter and vice versa. Pixels with less than 5 electron counts observed are plotted with grey.}
    \label{fig:model}
\end{figure}

Modelling the target pixel array is essential to the framework, not only for determining the flux fractions, but also for producing a model estimate of the centroid offset in the next step. It is therefore important that the model closely matches the observation. An example model pixel array is shown in the top plot of Figure \ref{fig:model} for TIC 279740441, with a comparison between the observation and model displayed in the bottom plot. The comparison is performed against the averaged observed target pixel array, which minimises the effect of photometric variability in the observation. The flux difference between their pixels is normalised by the shot noise of the averaged pixels in the observation. This allows the comparison plot to frame the difference in the context of the inherent Poisson noise. It should be noted that this $\sigma$ difference does not represent the error of the model pixels and is not used in that capacity.  For clarity, pixels with negative or less than 5 electron counts in the observed pixel array were removed from the comparison and plotted with grey, as small differences led to disproportionately high $\sigma$.

Analysing the comparison plot, we note that the model is generally in agreement with the observation, with the difference for all pixels below 3$\sigma$. As the method focuses on the flux contained in the aperture, the difference for these pixels is the only relevant metric for our purposes. To that effect, we computed the mean absolute $\sigma$ difference in the aperture pixels, which for the example in Figure \ref{fig:model} was found to be 1.21$\sigma$. We then repeated this process for 1000 randomly selected TIC stars and sectors in our dataset and taken the median of the differences. This resulted in a representative 1.37$\sigma$ difference for the method. This value suggests that the model is able to effectively reproduce the observation, at least in regards to the aperture. It also reveals that the shot noise is not enough to explain the difference between model and observation. This is not unexpected, as there are other sources contributing to the difference, including uncertainties in the interpolated PRF model and the flux calibration, the processing done on the observed flux, variability in the observation and unaccounted stars. To further quantify the difference we computed the mean absolute percentage difference for the aperture pixels, which was found to be 7.6\% for our example and 6.6\% for the randomly selected sample of stars.

Following the determination of the flux fractions, the implied eclipse depth, for each nearby source, $D_{ns}$, is calculated through the following relation: 

\begin{equation}
    D_{ns} = D_t \frac{f_t}{f_{ns}},
\end{equation}

where, $D_t$ is the measured depth of the event on the target and $f_t$ and $f_{ns}$ are the flux fractions for the target and the nearby source respectively. The implied depth is expressed as the fractional decrease of the observed normalised flux count. As the flux contribution of each source varies per sector, this process is repeated for all sectors, yielding a new set of fractions and implied depths. The mean eclipse depth of each source for all sectors is then calculated, with only sources with a mean depth below 1.0 considered as possible hosts of the event.

\subsection{Centroid Offset modelling}\label{Method:CentroidModel}
Calculating the implied eclipse depth allows us to model the expected centroid offset that would be produced if the event is on the target or on one of the possible nearby sources. First, the model PRF produced in the previous step is retrieved. As before, a model target pixel array is created, which corresponds to the out-of-transit observation. The out-of-transit centroid is then determined by computing the flux-weighted average horizontal and vertical pixel position as per Eq. \ref{eq: cent}.

Following, the expected in-transit centroid position is obtained for all possible sources of the event. This is produced by iteratively scaling the sources' flux count by their implied mean eclipse depth, which emulates the effect of the transit, and creating a new model target pixel array. This pixel array essentially represents the observation at the deepest point of the eclipse as it would occur on the possible source and allows us to calculate the in-transit centroid position. The difference between the model out-of-transit and in-transit centroid positions yields the model centroid offset vector, $\boldsymbol{\Vec{\Delta C_{m}}}$, as defined in Eq.\ref{eq: offset}. We impose a 10\% baseline error on both the vertical and horizontal components of the offset. This is higher than the average difference of 6.6\% found between the observed and modeled pixel arrays in Section \ref{Method:Depth}, to account for both outlier cases and the uncertainty of the implied eclipse depth. The suitability of this error is tested in Section \ref{Testing-Error}.

\subsection{Positional Probability}\label{Method:Probability}
The final step in the framework is the computation of the positional probability, which we define as the probability that a known stellar source is the host of the detected transit. The method relies on statistically comparing the model centroid offset, $\boldsymbol{\Vec{\Delta C_{m}}}$, to the observed, $\boldsymbol{\Vec{\Delta C_{obs}}}$, to derive the positional probability for each possible source of the event. The comparison utilises the Mahalanobis Distance \citep{mahalanobis, mahalanobis2}, \textit{d}, as follows:
\begin{equation} \label{eq: mahalanobis}
    d = \sqrt{(\boldsymbol{\Vec{\Delta C_{obs}}} - \boldsymbol{\Vec{\Delta C_{m}}})\boldsymbol{\Sigma}^{-1}(\boldsymbol{\Vec{\Delta C_{obs}}} - \boldsymbol{\Vec{\Delta C_{m}}})^T} ,
\end{equation}
where $\boldsymbol{\Sigma}^{-1}$ is the inverse covariance matrix. The covariance matrix is defined as:
\begin{equation} \label{eq: covmat}
    \boldsymbol{\Sigma} = \begin{bmatrix}\sigma_{x} & 0\\
                                         0 & \sigma_{y}
                            \end{bmatrix},
\end{equation} with:
\begin{equation} \label{eq: sigma}
    \sigma_{x} = \sqrt{\sigma_{\Delta C_{x_{obs}}}^2 + \sigma_{\Delta C_{x_{m}}}^2}, \sigma_{y} = \sqrt{\sigma_{\Delta C_{y_{obs}}}^2 + \sigma_{\Delta C_{y_{m}}}^2},
\end{equation}
    the respective uncertainties of the horizontal and vertical components of the observed and model centroid offsets added in quadrature. The derivation of these uncertainties is described in Sections \ref{Method:Centroid} and \ref{Method:CentroidModel} respectively.

The Mahalanobis distance can then be used to compute the likelihood of the source being the host of event from the following equation:
\begin{equation} \label{eq: likelihood}
    \mathcal{L}(s_{i}) = e^{\frac{-d^2}{2}},
\end{equation}
where $s_i$ denotes the $i^{th}$ source.

With the likelihood computed for all possible sources, we then compute the probability that the event is on a source using Bayesian inference:
\begin{equation} \label{eq: probability}
    \mathcal{P}(s_{i}|D,I) = \frac{\mathcal{L}(s_{i}) \mathcal{P}(s_{i}|I)}{\sum\limits_{i}\mathcal{L}(s_{i}) \mathcal{P}(s_{i}|I)}.
\end{equation}
As before, $s_i$ denotes the $i^{th}$ source, while $D$ and $I$ represent the observed data and the prior information respectively. $\mathcal{P}(s_{i}|I)$ is the prior probability for each source, which set as uniform for all possible sources. For the sources that do not meet the mean depth threshold described in Section \ref{Method:Depth}, the prior probability is essentially set to 0. As such, they are explicitly included in the framework, with a resulting positional probability of 0. Sources in the TIC with an object type other than a 'STAR' are also set to 0, as we confine the event only on stellar sources. It should be noted that the framework does not specifically exclude undetected stars as the hosts of the event, but rather constrains the probability on the known sources.

Finally, the process described in this section is implemented in a per-sector basis, with a positional probability derived only from the data of the sector. As the targets are observed in different TESS observation sectors, the process is repeated for each sector, resulting in a collection of probabilities for each source and sector. To combine the probabilities for the different sectors into one final probability for each source, we compute their median positional probability. We then re-normalize based on the sum of the median probabilities, so that the final probabilities sum up to 1. This approach ensures that we can produce standalone probabilities for each sector and allow for multi-sector observations to refine the final outcome. As a result, events observed in a varying number of sectors provide comparable results.

In the case of TOI-273.01, the centroid offset of which was used to showcase the method, the host source is the nearby star no. 5, TIC 279740439. The host was identified through follow-up observations by the TESS Follow-up Observing Program (TFOP) Working Group \citep{TFOP} and was reported in ExoFOP\footnote{https://exofop.ipac.caltech.edu/tess/target.php?id=279740441}. The \textit{SPOC} multi-sector image difference centroid analysis also pinpoint the location of the event to TIC 279740439. Our method provides a probability of 100\% for the host on both sector 34 and for all 7 sectors of observations combined. 

\begin{figure}
    \includegraphics[width=\columnwidth]{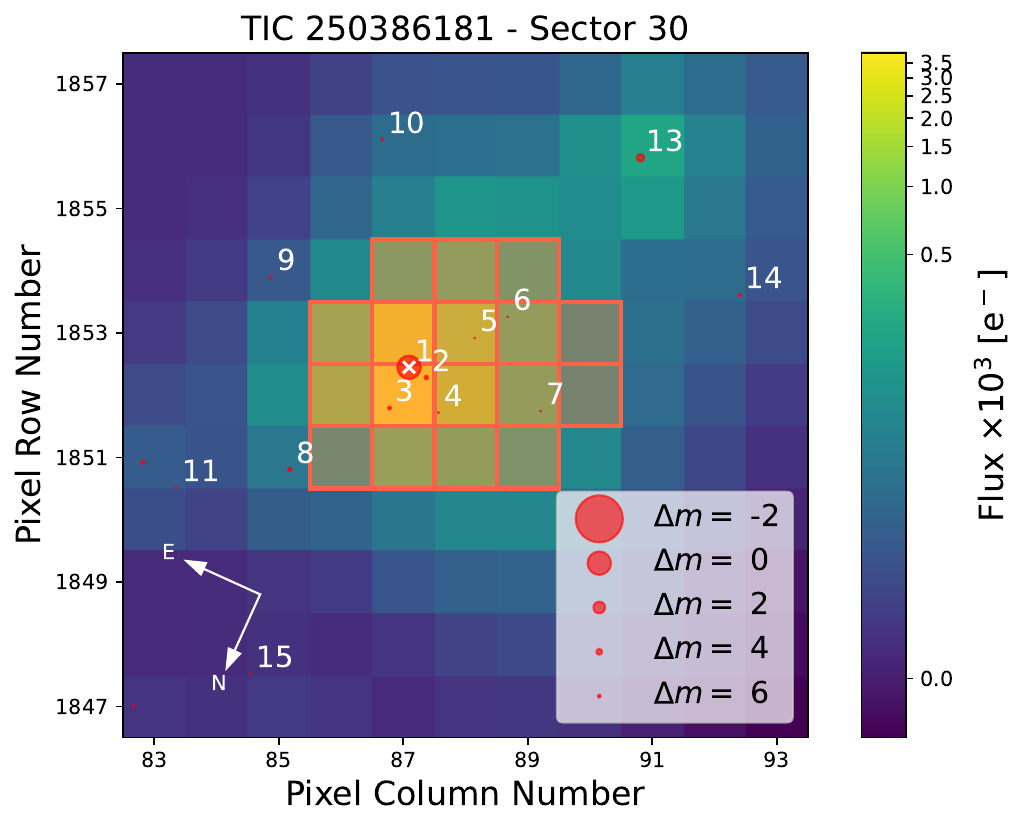}
    \caption{Observed \TESS\ target pixel array for TIC 250386181 in Sector 30, with the location for the star marked with an x. Resolved stellar sources are plotted as red circles with their size depending on their magnitude difference with the target. The shaded pixels represent the \textit{SPOC} aperture.}
    \label{fig:TP2}
\end{figure}

\begin{figure}
    \centering
    \includegraphics[width=\columnwidth]{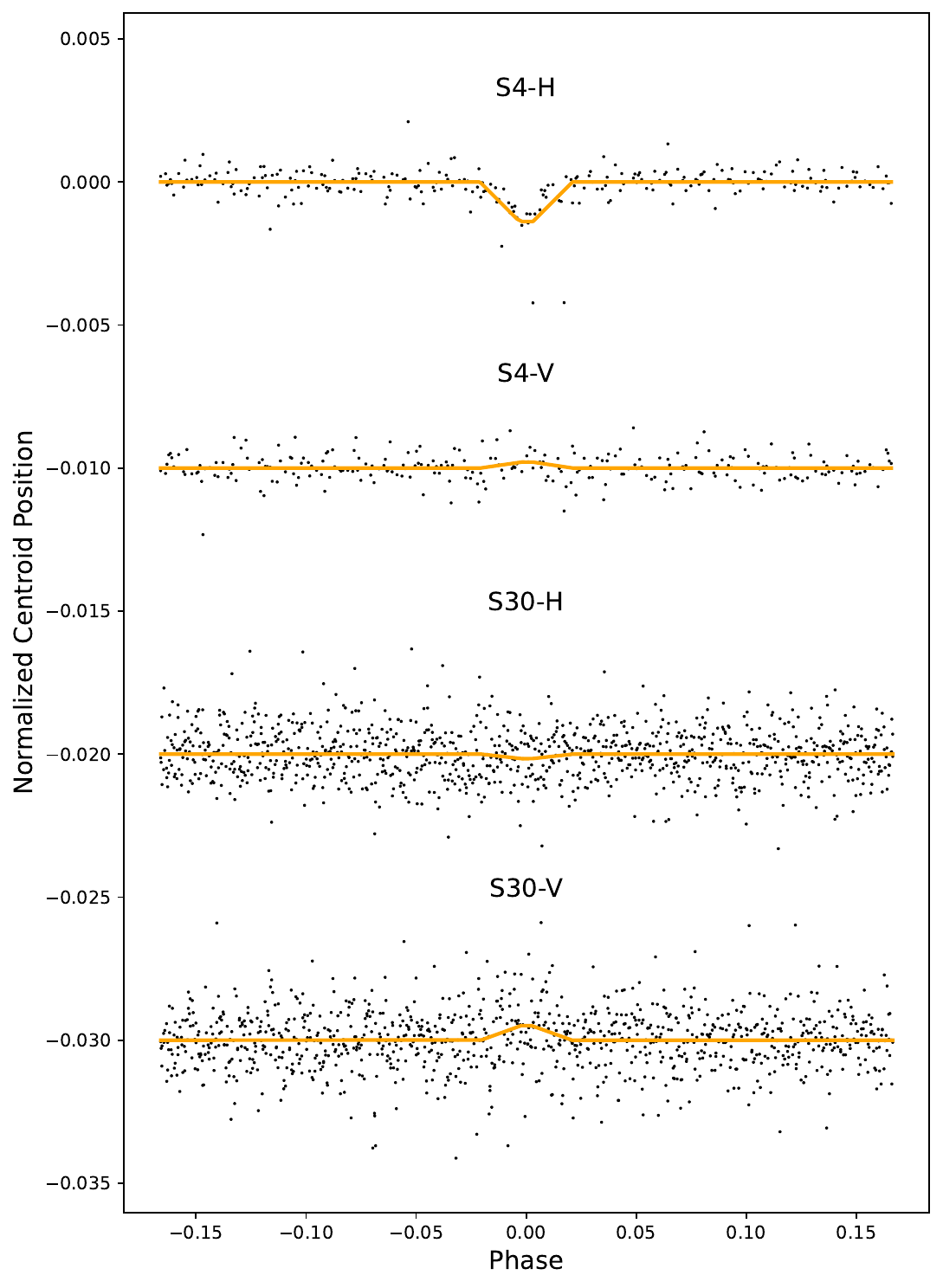}
    \caption{Normalized horizontal and vertical centroid positions for TIC 250386181 in Sectors 4 and 30, which have been offseted for clarity. The centroid positions have been phasefolded on the period of TOI-390.01, with the fitted trapezium model for each overlaid in orange.}
    \label{fig:centfit2}
\end{figure}

While TOI-273.01 is a case of a well separated host star, the algorithm performs effectively even on cases where the host star lies very close to the target. Such is the case of TOI-390.01 (TIC 250386181), which was observed in two \TESS\ sectors, 4 and 30. The target's pixel array for sector 30 is showcased in Figure \ref{fig:TP2}. The phasefolded horizontal and vertical centroids, along with their respective fits are displayed in Figure \ref{fig:centfit2}. The host star was identified as its close neighbour, TIC 250386182, labeled as no. 2 on the plot in Figure \ref{fig:TP2}, with a separation of 6.4$''$. For this case, the method attributed a probability of 79.7\% on the true host and 20.3\% on the target star.

\section{Testing} \label{sec:Testing}
\subsection{Overview} \label{Testing-Data}
The performance of the method was tested on known on-target and off-target TOIs. These included confirmed exoplanets, representing on target detections, and Nearby Eclipsing Binaries (NEBs) and Planet Candidates (NPCs), identified by TFOP. The list of NEBs and NPCs was compiled by examining the released observing notes of FP TOIs in ExoFOP\footnote{https://exofop.ipac.caltech.edu/tess/}. A recently released collection of on-target and off-target TOIs identified using \textit{Gaia} photometry \citep{GAIA_TESS} was also included in our testing. For all of the above categories, the only other requirement for their inclusion was the presence of \TESS\ FFI lightcurves released by \textit{SPOC}. These lightcurves are produced per sector from the FFI, with a sampling rate of 30 minutes for sectors 1-27 and 10 minutes for sectors 28-55 \citep{FFI}. For the purpose of these tests, only lightcurves up to sector 42 were used.

Overall, 3779 TOI were found, out of 5970 listed in the Exoplanet Archive at the time of our data selection, which had a released \textit{SPOC} FFI lightcurve up to sector 42. From those, we identified 691 known on-target and 176 off-target \TESS\ candidates, on which the method was tested. The results from these tests are presented in Sections \ref{Testing-PLNEB} and \ref{Testing-GAIA}. We further compiled a sample of 40 off-target events from the 176, for which the true host was already identified and listed in the observational notes on  ExoFOP. This afforded us the opportunity to test the method's ability in designating the likely true host and is detailed in Section \ref{Testing-NEB}. Additionally, we applied the method on 2408 of the TOIs which were designated as Planet Candidates, with the resulting probabilities presented in Section \ref{Testing-PC}. 

The transit data for the candidates were obtained from the NASA Exoplanet Arhcive\footnote{https://exoplanetarchive.ipac.caltech.edu/ Accessed:16/11/2022}, along with their TFOP disposition. For all events, the period was refined by maximising the significance of the transit signal over all available sectors. This was done due to the listed period in the archive often being determined when the event was initially detected in the early sectors. We purposefully avoided including a correction for doubling the period in our automated refinement process. Instead, we manually corrected some cases for which the observational notes in ExoFOP indicated that the period should be double. We expect that the effect of using half the period will have no impact on the method's outcome for events where the secondary eclipse has the same depth as the primary. For the rest of the events, we should still be able to obtain a centroid offset and possibly identify if the event is likely on-target or off-target, even though the resulting probabilities will not be entirely correct. 

Finally, before evaluating the method's performance, it is important to clarify our expectations. The method was developed as part of our upcoming validation framework, as a mean of quantifying the probability that a source is the host of the observed transit. The resulting probability is included in the prior of the validation framework and helps shape its outcome. The method was not designed to statistically confirm the source of the event, which is why we adopt a uniform prior for all possible source. Instead, it focuses on the ranking of the sources based on their probabilities, with the source that has the highest probability considered as the most likely host of the event. This probability ranking, in addition to shaping our validation prior, can help inform follow-up observation prioritisation. Therefore, it is important that the method is able to designate the right sources as the likely hosts of the event. The confirmation of the true host could then be potentially achieved through the follow-up observations or through the outcome of the validation framework.

Under this context, the outcome of the method for on-target events is considered successful if the target has the highest probability and is thus identified as the most likely host. On the other hand, for the off-target events, the method succeeds when the target does not have the highest probability. Ideally, the probability should tend towards 1 and 0 for the on-target and off-target events respectively, but that is of secondary importance. Finally, for off-target events where the true host is known, the outcome is considered successful if the true host has the highest probability. Even if the true host probability is below 0.5, which would suggest that the overall probability points to the event not being on that source, the outcome is still counted as a success if the host is ranked as the highest.

\subsection{Confirmed Planets and TFOP NEBs} \label{Testing-PLNEB}
\begin{figure}
    \centering
    \includegraphics[width=\columnwidth]{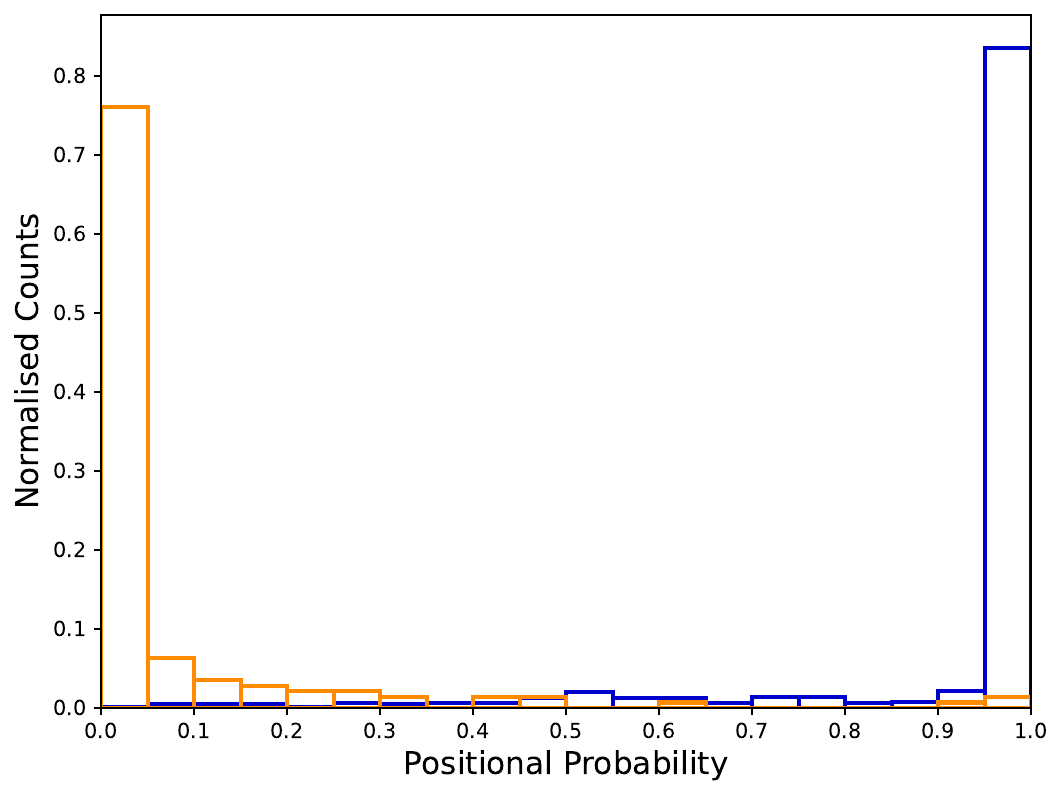}
    \caption{Positional Probabilities distribution for Confrimed Planets (blue) and NEBs (orange)}
    \label{fig:plvsneb}
\end{figure}

The first test focused on the positional probabilities of 658 TOIs listed in the Exoplanet Archive as Known and newly Confirmed \TESS\ Exoplanets. As the confirmation of their planetary nature is essentially a confirmation for an on-target detection, this group of TOIs was crucial for the development and evaluation of the method. Two of the planet TOIs were subsequently removed from the sample as they were found to have no visible transits on the available \textit{SPOC} FFI lightcurves. Additionally, the method was unable to produce positional probabilities for one TOI due to insufficient amount of centroid data following the processing of the centroid curve. Probabilities were successfully produced for the rest. 

This test also encompassed 144 off-target TOIs, of which 5 were classified as NPCs and the rest as NEBs. The targets were all identified through the efforts of TFOP's SG1 \citep{SG1} phase of follow-up observations. Positional probabilities were successfully computed for 142 of them, with one source failing due to poor centroid data quality and one removed as the event was not visible in the \textit{SPOC} FFI lightcurves. 

In order to assess the performance of the method on the two populations, the positional probabilities for the target sources of the two TOI populations were compared. This comparison is showcased in Figure \ref{fig:plvsneb}, with the two probability distributions plotted together. The resulting graph clearly demonstrates the method’s effectiveness in differentiating between on-target and off-target events. The distribution of confirmed planets, plotted in blue, is heavily skewed towards high probabilities, with the vast majority of the population having a positional probability higher than 0.95. This is the ideal outcome, as a high positional probability suggests a likely on-target event. Overall, 621 (94.8\%) planet TOI hosts had probabilities higher than 50\%, with 561 (85.6\%) laying above the 0.9 mark. For the off-target events, the distribution is similarly aligned with expectations. Of the 139 target sources, 117 (82.4\%) were found to have a positional probability less than 0.1 and 138 (97.2\%) less than or equal to 0.5. The dominance of low probabilities suggests that these events are likely off-target as expected.   

Only 34 planet TOI hosts had a positional probability less than 0.5, representing around 5\% of the population. While this suggests that there was a higher probability that the events were off-target rather than on-target, a comparison of the target to the rest of the nearby sources is required before identifying them as likely off-target. Therefore, we explored the difference in probability between the target and the most likely alternative source. In doing so, we found that 11 of those TOI hosts had a higher probability than their neighbours and as such were still considered to be the most likely source. This left only 23 problematic cases, leading to a failure rate of 3.5\%. On the known off-target cases, 5 TOI hosts were wrongly deemed by the method to be the most likely true host, representing 3.5\% of the population. The reasons that contributed to the above cases failing are discussed in Section \ref{Disc-Fail}. The overall number of problematic cases demonstrates that the method can indeed provide a good estimate on whether the target is the host of the TOI event.

\subsection{\textit{Gaia} Results} \label{Testing-GAIA}
A recently released collection of on and off-target TOIs identified through \textit{Gaia} photometry provided us with another opportunity to test the performance of the method. Although the list was comprised of 124 each of on-target and off-target TOIs, only 34 and 43 had \textit{SPOC} FFI lightcurves respectively. Positional probabilities were computed for all of them. The transits for one of the off-target TOIs, which was a QLP detection, were not visible in the SPOC FFI lightcurve and the event was subsequently removed from our sample.

\begin{figure}
    \centering
    \includegraphics[width=\columnwidth]{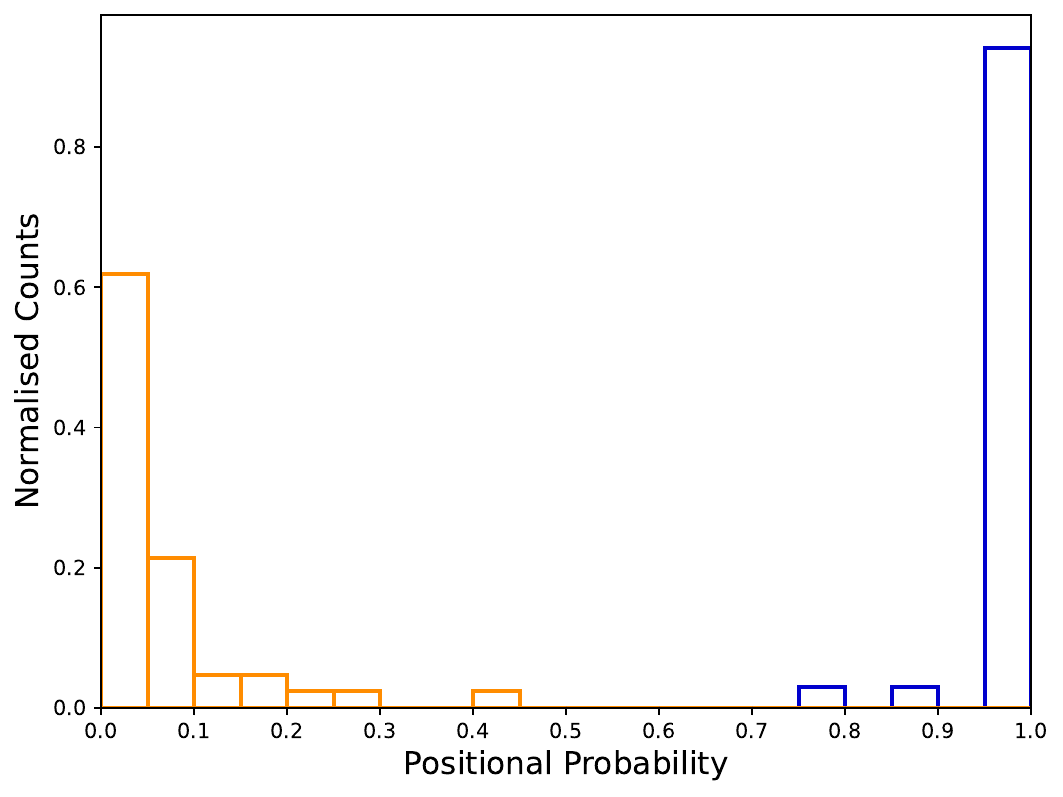}
    \caption{Positional Probabilities distribution for on-target (blue) and off-target (orange) TOIs identified with \textit{Gaia}.}
    \label{fig:GAIA}
\end{figure}

Similarly to the analysis for Section \ref{Testing-PLNEB}, the distribution of the positional probabilities for the target sources were plotted in Figure \ref{fig:GAIA}. The resulting distribution once again overwhelmingly proves that the method is effective in determining if the event is likely on the target or not. All but two of the on-target events had probabilities above 0.95, with the remaining two at 0.76 and 0.87 respectively, essentially designating them as on-target. At the same time, all of the off-target events had probabilities less than 0.5, which by itself would suggest that they are probably not occurring on the target sources. Examining the target sources against their nearby sources revealed just one target which was evaluated by the method as the most likely true host. This again proves a low failure rate for the method.    

\subsection{Comparison with known nearby true hosts} \label{Testing-NEB}
With the method proven to be effective at determining if the event is likely on or off target, the next step was to test if it could provide a reliable indication as to the true host for the event. To that effect, we explored the publicly available assessments from TFOP on ExoFOP, to find cases  of off-target TOIs where the true host was identified with a high degree of certainty. This yielded 38 NEB and 2 NPC cases with a known true host. We then examined the positional probabilities produced by the method for both the targets and their nearby sources, to determine the overall Nearby False Positive Probability (NFPP) for the event and if the true host had the highest positional probability. The NFPP is derived as the difference from unity of the target's positional probability and therefore expresses the overall probability of the event being off-target. A nearby star with a positional probability equal to the NFPP would thus be the sole likely alternative source of the event. Ideally then, the probability of a true host would not only be the highest, but also equal to the NFPP. 

The results of this test, sorted by ascending TOI number, are listed in Table \ref{tab: SG1_Comparison}. As it can be surmised, the method is effective in indicating which star is the likely true source of the transit. For all but 2 of the TOIs the identified true hosts had the highest positional probabilities and were thus deemed by the method as the most likely sources of their respective events. In addition, for the majority of cases the true hosts had positional probabilities close to the overall NFPP and all events had a high NFPP, with the lowest being 68.8\%. In regards to the two failure cases, the true host for TOI 531.01 was the second most likely alternative source. However, a brighter star in the same direction but closer to the target had an expected centroid shift that more closely matched the observation, resulting to a much higher probability. The true host for TOI 1256.01 was not selected as a possible alternative source, most likely due to being a bright star at a large distance to the target (102.26$''$) in a very crowded area. This is a challenging case and suggests a potential limitation of the method. 

Overall, this is considered a successful test for the framework and demonstrates the method's capability in indicating the true host of the event. Despite its small size, the sample was found to be representative of the larger TOI population as a whole, with a median Signal-to-Noise Ratio (SNR) of 16.7 for the sample and 20.9 for the 3186 TOIs used in the rest of our tests. Therefore, the results from this test present an excellent argument for employing the method to identify and assess the possible alternative sources of TESS transit events.

\begin{table*}
    \begin{threeparttable}[b]
    \centering
    \caption{The Nearby False Positive Probabilities (NFPP) and true host probabilities for NEB and NPC TOIs where the true host has been identified by TFOP. The last column specifies whether the true host was ranked as the highest possible source of the event.}
    \label{tab: SG1_Comparison}
        \begin{tabular}{ccccccc}
        \hline
        \hline
            \textbf{Target} & \textbf{TOI} & \textbf{TFOP Disposition} & \textbf{True Host} & \textbf{NFPP} & \textbf{True Host Probability} & \textbf{True Host Highest} \\ 
            \hline
            403287048 & 152.01 & NEB & 403287050 & 0.9645 & 0.9175 & Y \\ 
            260043723 & 217.01 & NEB & 260043722 & 0.9997 & 0.7634 & Y \\ 
            279740441 & 273.01 & NEB & 279740439 & 1.0000 & 1.0000 & Y \\ 
            139285736 & 377.01 & NPC & 139285741 & 1.0000 & 1.0000 & Y \\ 
            220396259 & 379.01 & NEB & 220396256 & 1.0000 & 1.0000 & Y \\ 
            167418898 & 383.01 & NPC & 167418903 & 0.6688 & 0.6143 & Y \\ 
            92359850 & 387.01 & NEB & 92359852\tnote{*} & 1.0000 & 0.9965 & Y \\ 
            250386181 & 390.01 & NEB & 250386182 & 0.7972 & 0.7968 & Y \\ 
            219388773 & 399.01 & NEB & 219388775 & 0.9998 & 0.9664 & Y \\ 
            176778112 & 408.01 & NEB & 176778114 & 1.0000 & 1.0000 & Y \\ 
            279251651 & 419.01 & NEB & 766100791 & 1.0000 & 0.9877 & Y \\ 
            427352241 & 485.01 & NEB & 427352247 & 1.0000 & 0.9999 & Y \\ 
            108645766 & 497.01 & NEB & 108645800 & 0.9544 & 0.2514 & Y \\ 
            274138511 & 506.01 & NEB & 760244235 & 0.9935 & 0.3609 & Y \\ 
            431999925 & 513.01 & NEB & 431999916 & 0.9996 & 0.8818 & Y \\ 
            438490744 & 529.01 & NEB & 438490748 & 1.0000 & 1.0000 & Y \\ 
            302895996 & 531.01 & NEB & 302895984 & 1.0000 & 0.1329 & N \\ 
            59003115 & 556.01 & NEB & 59003118 & 0.7315 & 0.6269 & Y \\ 
            1133072 & 566.01 & NEB & 830310300 & 1.0000 & 0.9999 & Y \\ 
            146463781 & 636.01 & NEB & 146463868 & 1.0000 & 0.6887 & Y \\ 
            432008938 & 643.01 & NEB & 432008934 & 1.0000 & 0.9257 & Y \\ 
            54085154 & 662.01 & NEB & 54085149 & 0.9199 & 0.7789 & Y \\ 
            373424049 & 742.01 & NEB & 373424060 & 0.8024 & 0.4892 & Y \\ 
            271596418 & 868.01 & NEB & 271596416 & 0.9221 & 0.2785 & Y \\ 
            364107753 & 909.01 & NEB & 1310226289 & 0.8155 & 0.4457 & Y \\ 
            253990973 & 1061.01 & NEB & 253985122 & 1.0000 & 0.4945 & Y \\ 
            375225453 & 1096.01 & NEB & 375225447 & 0.9996 & 0.6484 & Y \\ 
            117256577 & 1156.01 & NEB & 117256550 & 1.0000 & 1.0000 & Y \\ 
            30329585 & 1175.01 & NEB & 30329564 & 1.0000 & 0.9829 & Y \\ 
            237185205 & 1191.01 & NEB & 237194795 & 1.0000 & 1.0000 & Y \\ 
            274762761 & 1256.01 & NEB & 274762865 & 1.0000 & 0.0000 & N \\ 
            267561446 & 1284.01 & NEB & 267561450 & 0.9814 & 0.7154 & Y \\ 
            274662200 & 1285.01 & NEB & 274662220 & 0.9998 & 0.4641 & Y \\ 
            408203470 & 1289.01 & NEB & 408203452 & 1.0000 & 0.6298 & Y \\ 
            233681149 & 1340.01 & NEB & 233681148 & 0.9998 & 0.9995 & Y \\ 
            161156159 & 1988.01 & NEB & 161156158 & 1.0000 & 1.0000 & Y \\ 
            315755496 & 2053.01 & NEB & 315755494 & 0.7713 & 0.7713 & Y \\ 
            258402425 & 2156.01 & NEB & 258402422 & 0.9982 & 0.5412 & Y \\ 
            284564230 & 2162.01 & NEB & 284564232 & 0.8709 & 0.7391 & Y \\ 
            230075120 & 4110.01 & NEB & 230075121 & 0.8899 & 0.8119 & Y \\  
            \hline
        \end{tabular}
        \begin{tablenotes}
            \item [*] TIC 92359852 was later found to correspond to two \textit{Gaia} sources, with TIC IDs 651667036 and 651667038. The probability presented here is the sum of the probabilities for the two sources.
        \end{tablenotes}
    \end{threeparttable}
\end{table*}

\subsection{Model Offset Error} \label{Testing-Error}
The list of known off-target TOI hosts, along with all the on-target detections, provide us with the opportunity to examine the difference between the method's model centroid offset and the observed, especially in relation to their uncertainties. As the host for all these cases is known and the depth can either be directly measured or extrapolated, the expectation is that offset should be accurately modeled and match the observation within error. Moreover, the error itself should be Gaussian in nature, with the difference expected to be mostly driven by the random fluctuations of the flux in the observations. 

\begin{figure}
    \centering
    \includegraphics[width=\columnwidth]{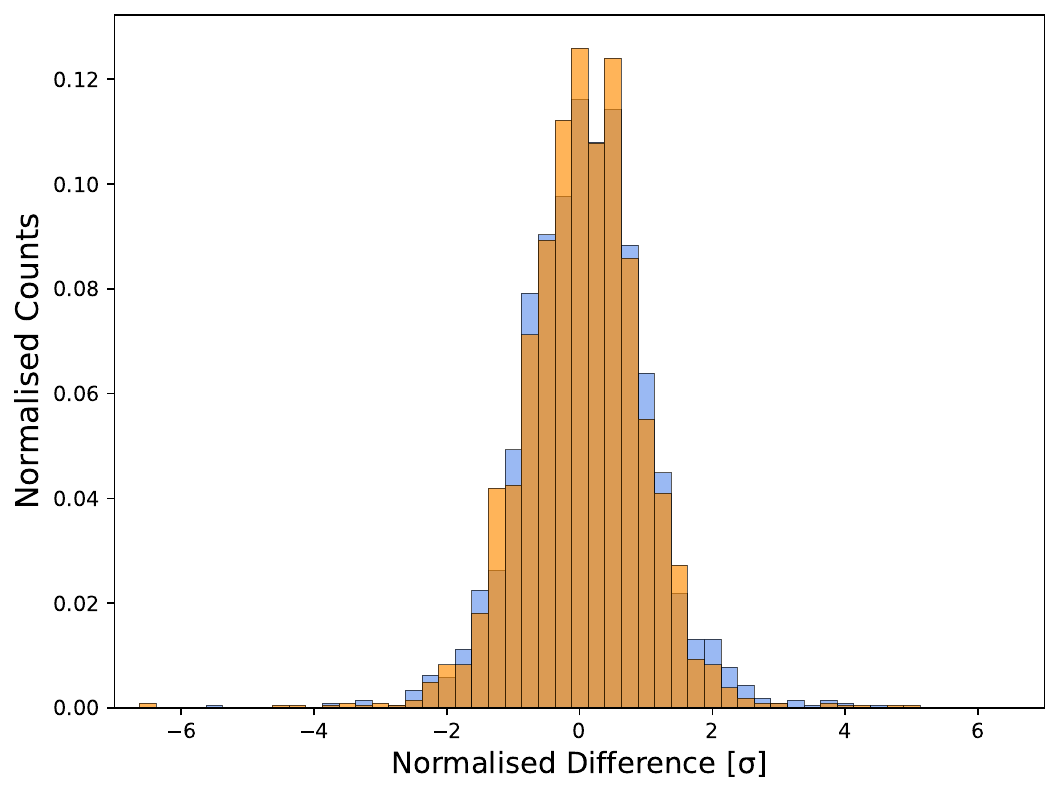}
    \caption{Distribution of the normalised differences between the horizontal (blue) and vertical (orange) components of the observed and modelled centroid offsets for known TOI hosts. The difference is normalised by the propagated uncertainties of the observed and modelled offset for each axis.}
    \label{fig:Error}
\end{figure}

To test this, we computed the difference between the horizontal and vertical components of the modelled and observed offsets for each sector of observation and normalised it by their propagated uncertainties. These uncertainties were computed as described in Section \ref{Method:Centroid} for the observed offset and Section \ref{Method:CentroidModel} for the model offset. 
The resulting distributions of the normalised differences is plotted in fig. \ref{fig:Error} and encompass the differences for 728 TOI, with a total of 2117 observations. As seen in the plot, the distribution for both the horizontal and vertical components appears Gaussian, with a mean and standard deviation of (-0.02, 0.99) and (-0.08, 0.94) respectively. This is in support of our hypothesis as stated above and suggests that the uncertainties for both the observed and the model offset are well defined. For the model centroid offset especially, this result supports our decision to adopt a baseline 10\% error.

\subsection{Planet Candidates} \label{Testing-PC}
\begin{figure}
    \centering
    \includegraphics[width=\columnwidth]{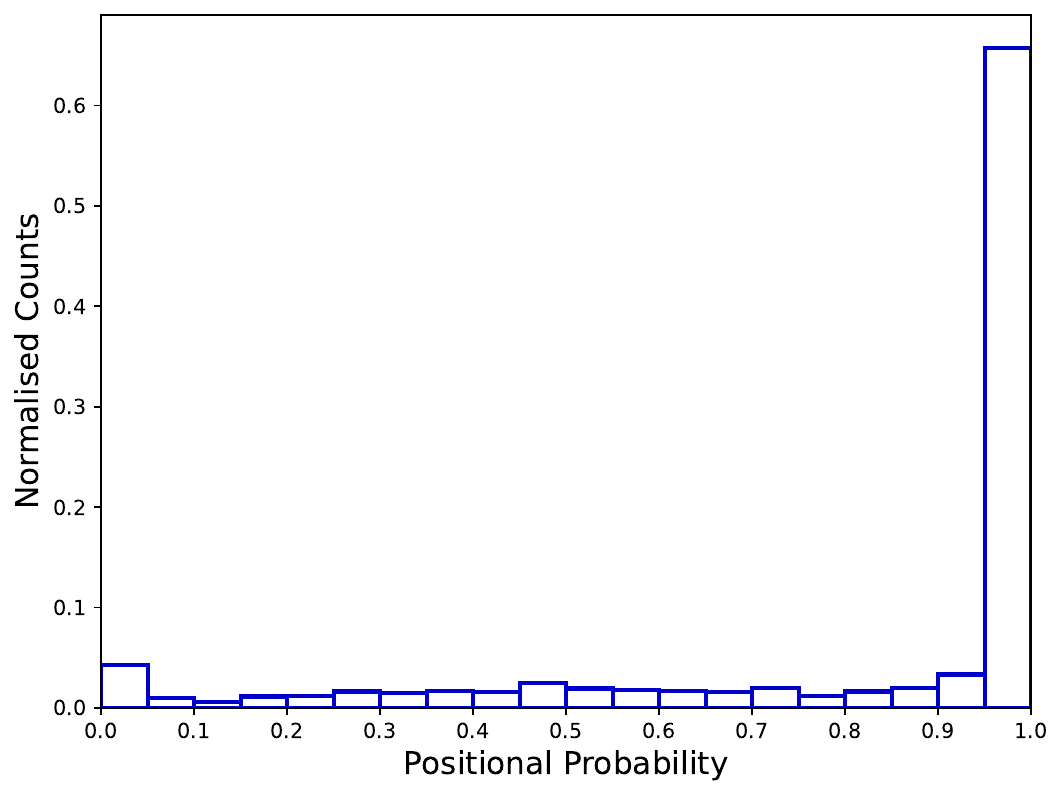}
    \caption{Positional Probabilities distribution for active planet candidate TOIs.}
    \label{fig:PC}
\end{figure}

Finally, we applied the method on 2408 TOIs which were listed a Planet Candidates, with positional probabilities successfully produced for 2365 of them. The distribution of the probabilities is displayed in Figure \ref{fig:PC}. These are all targets where the location of the event has not been yet identified through any follow-up observations. Therefore, this is the first complete set of event-location information for this set of TOIs. What can be readily surmised by the graph is that the majority of the detections are indeed on target events, with a sizeable population of off-target events. In more detail, applying similar analysis as for the known on and off target TOIs, 2072 candidates were designated by the method as most likely on target and 293 as most likely off target. The full list of positional probabilities for these TOI events, which includes both the target stars and nearby sources with probability higher than 0.01\%, can be found in Table \ref{tab: PC}.

\begin{table}
    \centering
    \caption{Probabilities for target stars and nearby alternative sources of PC TOIs. Targets are identified by TIC ID. Full table is available online.}
    \begin{tabular}{cccc}
        \hline
        \hline
        \textbf{TOI} & \textbf{Assumed Target} & \textbf{Source} & \textbf{Source Probability} \\
        \hline 
        119.01 & 278683844 & 278683844 & 1.0000 \\ 
        119.02 & 278683844 & 278683844 & 0.9995 \\ 
        119.02 & 278683844 & 278683840 & 0.0005 \\ 
        121.01 & 207081058 & 207081058 & 1.0000 \\ 
        124.01 & 29831208 & 29831208 & 1.0000 \\ 
        128.01 & 391949880 & 391949880 & 0.5097 \\ 
        128.01 & 391949880 & 675054894 & 0.4660 \\ 
        128.01 & 391949880 & 391949878 & 0.0169 \\ 
        128.01 & 391949880 & 391949876 & 0.0067 \\ 
        128.01 & 391949880 & 391949884 & 0.0007 \\ 
        131.01 & 235037761 & 235037761 & 1.0000 \\ 
        \vdots & \vdots & \vdots & \vdots \\
        \hline
    \end{tabular}
    \label{tab: PC}
\end{table}

\section{Discussion}\label{sec:Discussion}

\subsection{Effectiveness and Usability} \label{Disc-Effectiveness}
Testing the method successfully showcased its effectiveness, which more than exceeded our expectations. The method was able to assign appropriate probabilities for 688 known on-target events and 173 off-target events in 96.6\% of the cases. This high success rate suggests that the method can be reliably used to quickly test the locality of a detected \TESS\ event. Even more so, it can provide an indication as to the most likely source of an event in the case of an off-target detection, with the test revealing a 95\% success rate in designating already known true hosts as the most likely sources of the events. 

Moreover, this is achieved with a fast runtime and minimal resources. The \TESS\ \textit{SPOC} lightcurves for the target, PRF models and access to the TIC are all that are needed. As a result, an assessment for a single target can be performed within a minute and the process is easily parallelisable. This can be extremely beneficent, especially when taking into account the large number of TOIs. Follow-up observations and clearing the field of nearby sources is both a time and resource consuming process. The tool can potentially alleviate the need to wait for host identification part of the follow-up process to be completed, by providing an assessment as to the suitability of the event for further research. In addition, it can augment the follow-up process, by identifying the most likely alternative hosts that can then be prioritised for observation. It should be noted that the above performance and resources usage are for the \textit{SPOC} lightcurves, for which this method was developed, as they include measurements of the photometric centroid. If the measurements are not included, then the target pixel file would be needed as well, with additional processing required to obtain the centroid curves. At present the method can only be used for the \textit{SPOC} lightcurves. However, we plan on modifying the method in the future it to enable its use on lightcurves from other pipelines as well.

Furthermore, the probabilities obtained by this method can be used as further diagnostics to those provided by the SPOC and QLP pipelines for detected TOIs. They can also be incorporated in vetting and validation pipelines. This will be the case for our upcoming \textit{RAVEN} validation pipeline, the paper of which is in preparation. 

\subsection{Comparison with other methods} \label{Disc-Comp}
Our method expands upon the photometric centroid offset described in \citet{Centroid}. Instead of using the observed offset to determine the location of the event, we are modelling and comparing the centroid that would be observed if the event was on different sources, including the target. Therefore, we directly query the true host of the event by finding the model that best reproduces the observation, rather than infer it based on the event's location. As mentioned in \citet{Centroid}, the determination of the event location using the photometric centroid offset can be susceptible to other sources of variability. The use of a generative model allows us to minimise the effect of photometric variability and to the incorporate additional observational data into the computation, such as the transit depth and the magnitude and location of the sources. 

Compared to the imaging difference method, our tool offers an alternative for assessing centroid offsets. As described in the introduction, the imaging difference method identifies the pixel with the highest flux change during the transit cadences. A PRF fit to the image difference is then used to identify the location of the source of the change. This location will most of the time point to a single star, the likely source of the transits, especially when determined over many sectors of observations and averaged. However, the location can sometimes point to no particular source and/or be consistent with a few nearby stars that fall within the 3$\sigma$ threshold often adopted for the method. An example for each of the above cases can be seen in the multi-sector data validation reports produced by the SPOC pipeline for TOI-273.01 (TIC 279740441) and TOI-1363.01 (TIC 230017325), which are available on the MAST Portal \footnote{https://mast.stsci.edu/portal/Mashup/Clients/Mast/Portal.html}. As already stated in the comparison with the photometric centroid above, our method is querying directly the true host of the transit, instead of determining its location. By confining the source of an observed transit to known stars and  producing probability estimates based on generative modelling, we quantify the likelihood of each to be the host the detected event. This allows for comparing and ranking each source, which can be beneficial in both identifying the most likely true source and for follow-up prioritisation of the nearby sources. The two methods therefore offer independent and complementary results and can be used together to enhance the reliability of assessing the true source of the detected TESS transits.

The most comparable method to the one presented here is the \citet{Bryson2017}. The method was developed for Kepler Objects of Interest and aimed to improve upon 3$\sigma$ threshold approach for the Imaging Difference technique, by producing the positional probability of the event being on the target or the nearby known stars. At its core, the method relies on modeling the imaging difference that would be produced if the host of the event was one of the known stars in the Kepler Input Catalog and computing its likelihood ratio. The positional probability is then determined using a likelihood ratio. Our method is conceptually similar to the \citet{Bryson2017} method, although the implementation differs, with one key difference being the use of the photometric centroid instead of the imaging difference technique. A direct comparison between the performance of the two methods cannot be made, as they are employed for different missions.

Any comparison with the validation tools is not straightforward. The method does not perform any analysis into the nature of the event. It presents a result purely focused on the observed transit and centroid offset. If identifying the true host of the event is the desired outcome, the validation methods do not offer this functionality. Only, triceratops is accounting for the known nearby sources when examining the nearby false positive scenarios, but it does not directly determine the true host. Our method aims to distinguish the true host, but not statistically confirm it. It also does not provide the probability of a FP detection, which is what the validation methods are designed to do. Therefore, a direct comparison will need to be done based on the in-development \textit{RAVEN} validation pipeline, for which the method presented here is a major part.

\subsection{Limitations} \label{Disc-Fail}

For about 3.4\% of the cases, for both on-target and off-target TOIs, the method produced positional probabilities that were not consistent with expectations. The main reasons that contributed to these "failure" cases are discussed here.

\begin{figure}
    \centering
    \includegraphics[width=\columnwidth]{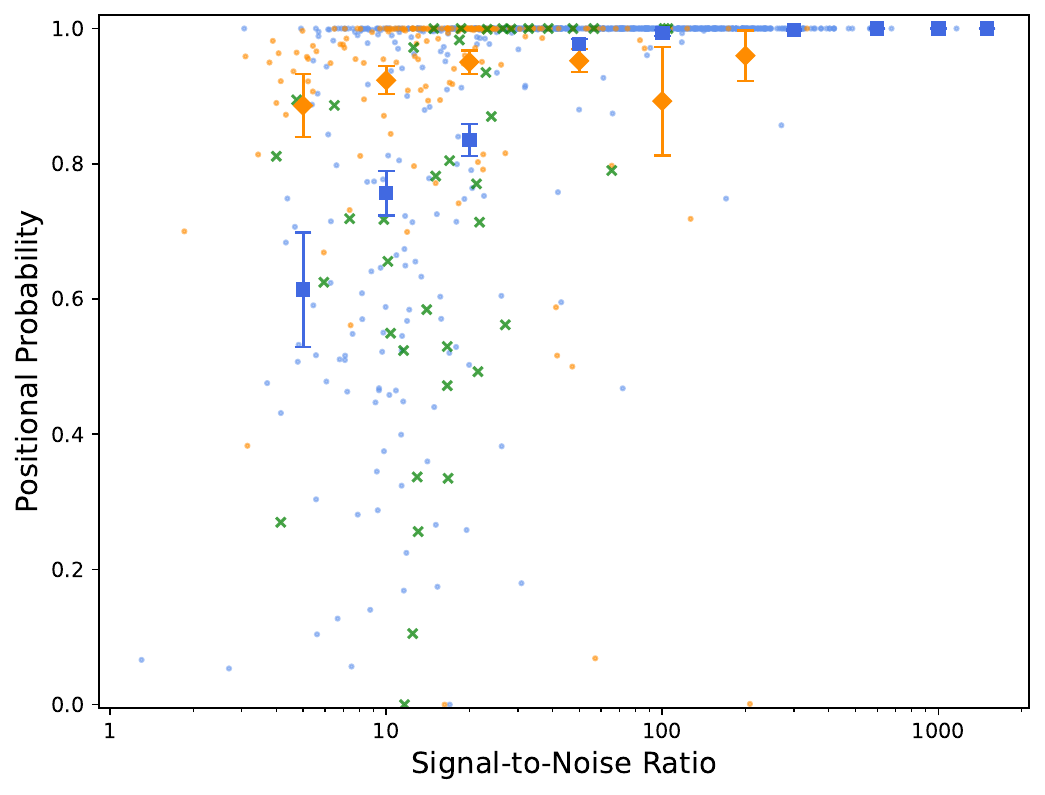}
    \caption{The positional probability for on-target (blue) and off-target (orange) TOI in relation to the median Signal-to-Noise Ratio (SNR) of the detected transit over the different sectors of observation. For the off-target events, the probability that the event is not on the target is displayed. The square and diamond markers represent the mean probability of SNR bins for on-target and off-target TOI events respectively. The green markers display the probability of the known off-target true hosts listed in Table \ref{tab: SG1_Comparison}.}
    \label{fig:SNR}
\end{figure}

Low SNR transits: these transits can often result in multiple contaminating sources being considered as possible hosts of the event, depending on their flux contribution. Moreover, they usually feature small, noise dominated, observed centroid offsets, especially in the case of on-target events. This can lead to at least a few of the possible hosts having model offsets that are statistically compatible with the observation. As a result, the final probability would become more diluted due to the increase in the number of compatible hosts. The effect of the transit SNR on the positional probability is explored in figure \ref{fig:SNR}, which displays the probability of known on-target (blue) and off-target (orange) TOIs in relation to the transit SNR. The probabilities of the 40 known true hosts for the off-target cases in Table \ref{tab: SG1_Comparison} are also plotted with green, to provide additional context. As can be seen in the figure, the on-target probability appears to be negatively impacted by the lower SNR. This is not surprising, as the diluted probability lowers the target's probability. For the off-target probability, there is only a tentative suggestion of lower performance at low SNR values, which could be attributed to the target having a non-negligible probability due to the uncertainty in the small centroid offsets. However, the overall off-target probability is still high and the impact of the transit SNR appears to be minimal. For the off-target hosts, their small number does not allow us to infer any definitive correlation between the positional probability and the SNR. However, their scatter seems to resemble that of the on-target cases, suggesting a similar performance.

\begin{figure}
    \centering
    \includegraphics[width=\columnwidth]{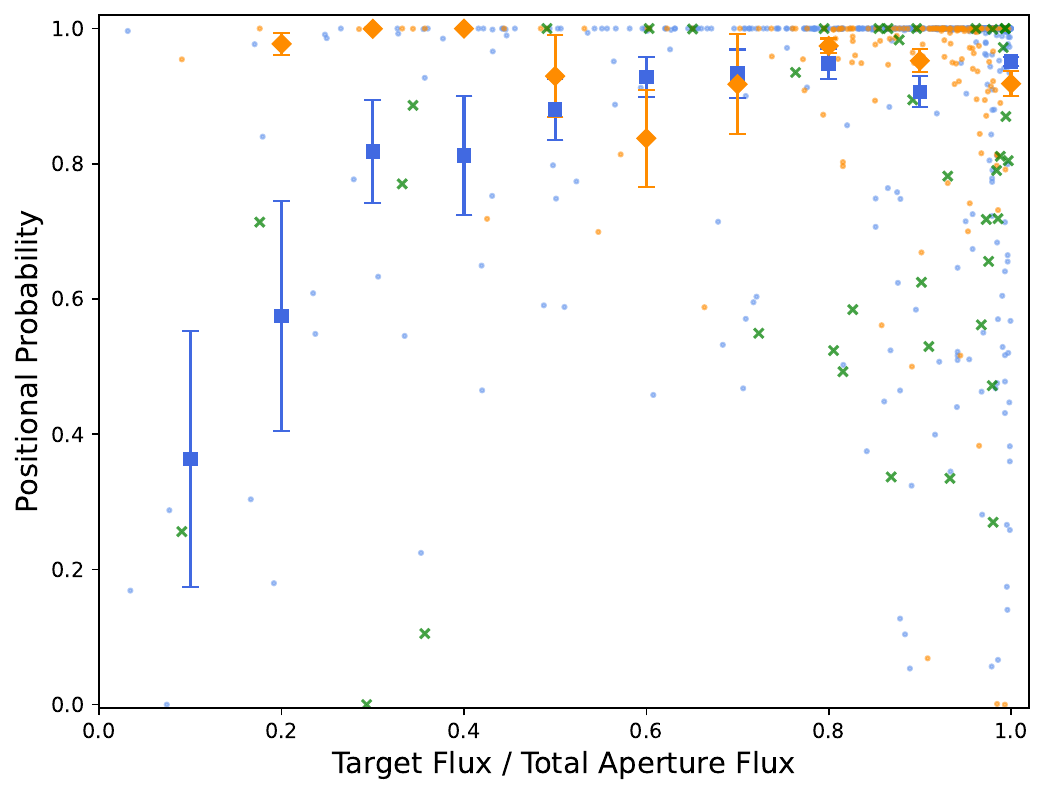}
    \caption{The positional probability for on-target (blue) and off-target (orange) TOI in relation to the fraction of flux from the target star in the aperture, as reported by the SPOC pipeline. The fraction presented in the figure is the median for the target over the different sectors of observation. For the off-target events, the probability that the event is not on the target is displayed. The square and diamond markers represent the mean probability of crowding fraction bins for on-target and off-target TOI respectively. The green markers display the probability of the known off-target true hosts listed in Table \ref{tab: SG1_Comparison}.}
    \label{fig:Crowding}
\end{figure}

Crowded Target Pixel Arrays: these refer to observations in regions with high density of stars. Naturally there is a greater degree of blending in the aperture for these observations and thus more possible hosts for the detected transit. Moreover, modelling the crowded target pixel array was found to present a greater challenge, leading to higher than average differences, which could affect the estimations for the flux fractions and the centroid offset. We explore the effect of crowding in figure \ref{fig:Crowding}, using the fraction of the flux in the \textit{SPOC} aperture that is originating from the target star, as reported by the pipeline, to represent it. Based on the graph, the probability for the on-target TOI events decreases as the degree of crowding increases, suggesting a possible correlation between the two. However, it is evident that very few on-target TOIs occupy this low-fraction, high-crowding space. The majority is lying above 0.8, where there is a high degree of scattering for the probability. As a result, no definitive assessment can be made on the correlation between the on-target probability and crowding. This is also true for both the off-target probability and for the probability of the known off-target true hosts.

Model uncertainties: Both the PRF interpolation process and any uncertainty in the \TESS\ magnitude can result to model observations with greater difference than average to the \TESS\ observations. This can again result to incorrect flux fractions. For 2 of the 5 problematic NEBs in Section \ref{Testing-PLNEB}, the flux fractions, combined with the recorded depths led to no possible alternative sources for the event. 

Host star in a visual Binary: this affects known and confirmed planets orbiting one of the binary stars. The differences in the expected centroid offsets from the two sources are usually sufficiently small, resulting in an ambiguous location determination between the two. This was seen in 3 of the problematic planet cases in Section \ref{Testing-PLNEB}.

\section{Conclusion}\label{sec:Conclusion}
In conclusion, we present a method to assess the host of a detected \TESS\ event, based on the observed transit and its associated photometric centroid offset. The method produces good probability estimates as to the likelihood of the event being on the target or any possible nearby sources, evident by the positional probabilities obtained for known on-target and off-target events. Moreover, the method was shown to also be effective in providing reliable indications as to the true source of the transit in the majority of tested cases. It does so with a quick runtime and minimal additional resources. It can therefore serve to augment the current efforts to determine the true host of a detected TOI, which include centroid assessments by the \textit{SPOC} and QLP pipelines and the SG1 follow-up phase of TFOP. It can be run almost immediately following a TOI detection and can act as another test for the validity of the event. 

The method was designed to run with \textit{SPOC} FFI lightcurves, but can also be used on 2-minute cadence lightcurves. It will be upgraded in the future, so that it can also run on QLP lightcurves. The method will form a core part of an in-development vetting and validation pipeline, but is released at the current time as a standalone tool. 

\section*{Acknowledgements}
We thank the anonymous referee, whose insightful review helped improve the quality of this paper.

This paper includes data collected by the \TESS\ mission. Funding for the \TESS\ mission is provided by the NASA Explorer Program. Resources supporting this work were provided by the NASA High-End Computing (HEC) Program through the NASA Advanced Supercomputing (NAS) Division at Ames Research Center for the production of the \textit{SPOC} data products.

We acknowledge the use of public \TESS\ Alert data from pipelines at the \TESS\ Science Office and at the \TESS\ Science Processing Operations Center.

This research has made use of the Exoplanet Follow-up Observation Program (ExoFOP; DOI: 10.26134/ExoFOP5) website, which is operated by the California Institute of Technology, under contract with the National Aeronautics and Space Administration under the Exoplanet Exploration Program.

This work made use of \texttt{tpfplotter} by J. Lillo-Box (publicly available in www.github.com/jlillo/tpfplotter), which also made use of the python packages \texttt{astropy}, \texttt{lightkurve}, \texttt{matplotlib} and \texttt{numpy}.

AH is supported by an STFC studentship. 
DJA is supported by UKRI through the STFC (ST/R00384X/1) and EPSRC (EP/X027562/1).

\section*{Data Availability}
This paper exclusively uses data from the \textit{TESS} mission. All data from the mission is available to the public via the Mikulski Archive for Space Telescopes (MAST)\footnote{https://archive.stsci.edu/missions-and-data/tess}.




\bibliographystyle{mnras}
\bibliography{ref} 


\bsp	
\label{lastpage}
\end{document}